\documentclass[prb,twocolumn,showpacs,preprintnumbers,amsmath,amssymb,floatfix]{revtex4-1}

\usepackage{graphicx,color,hyperref}

\begin{document}

\title{Trail-Needs pseudopotentials in quantum Monte Carlo calculations with
  plane-wave/blip basis sets}

\author{N.\ D.\ Drummond}

\affiliation{Department of Physics, Lancaster University, Lancaster LA1 4YB,
  United Kingdom}

\author{J.\ R.\ Trail}

\affiliation{TCM Group, Cavendish Laboratory, University of Cambridge, 19
  J.\ J.\ Thomson Avenue, Cambridge CB3 0HE, United Kingdom}

\author{R.\ J.\ Needs}

\affiliation{TCM Group, Cavendish Laboratory, University of Cambridge, 19
  J.\ J.\ Thomson Avenue, Cambridge CB3 0HE, United Kingdom}

\begin{abstract}
We report a systematic analysis of the performance of a widely used set of
Dirac-Fock pseudopotentials for quantum Monte Carlo (QMC) calculations.  We
study each atom in the periodic table from hydrogen ($Z=1$) to mercury
($Z=80$), with the exception of the $4f$ elements ($57 \leq Z \leq 70$).  We
demonstrate that ghost states are a potentially serious problem when
plane-wave basis sets are used in density functional theory (DFT)
orbital-generation calculations, but that this problem can be almost entirely
eliminated by choosing the $s$ channel to be local in the DFT calculation; the
$d$ channel can then be chosen to be local in subsequent QMC calculations,
which generally leads to more accurate results.  We investigate the achievable
energy variance per electron with different levels of trial wave function and
we determine appropriate plane-wave cutoff energies for DFT calculations for
each pseudopotential.  We demonstrate that the so-called ``T-move'' scheme in
diffusion Monte Carlo is essential for many elements.  We investigate the
optimal choice of spherical integration rule for pseudopotential projectors in
QMC calculations.  The information reported here will prove crucial in the
planning and execution of QMC projects involving beyond-first-row elements.
\end{abstract}


\maketitle

\section{Introduction}

Quantum Monte Carlo (QMC) methods are highly accurate, explicitly correlated
wave-function-based techniques for calculating material properties from
first-principles.  The diffusion quantum Monte Carlo (DMC)
method\cite{Ceperley_1980,Foulkes_2001} is generally regarded as the most
accurate first-principles method available for studying condensed matter.
However, one significant weakness of DMC is its poor scaling with atomic
number $Z$ in all-electron calculations.  The computational expense of
achieving a given statistical error bar scales as
$O(Z^5)$--$O(Z^{6.5})$.\cite{Hammond_1994,Ceperley_1986,Ma_2005} The physical
origin of the problem is that the short length scale associated with core
electrons (the Bohr radius of the hydrogen-like ion, which goes as $1/Z$)
results in the need for very small time steps at large $Z$; on top of this
there is the obvious additional expense due to the increase in the number of
electrons with $Z$.  In practice, it is usually only feasible to perform
all-electron DMC calculations for first-row atoms.  Furthermore, performing
all-electron DMC calculations for anything beyond lithium generally requires
the use of Gaussian or Slater-type basis sets rather than plane-wave basis
sets to represent orbitals.  For these reasons, the great majority of DMC
studies of condensed matter have used pseudopotentials to represent atomic
cores.\cite{Needs_2010}

The use of nonlocal pseudopotentials introduces new difficulties, however.
One issue is that the ``standard'' DMC algorithm assumes the potential-energy
operator to be local.  The most widely used solution to this difficulty is the
pseudopotential locality approximation,\cite{Hurley_1987} in which the
nonlocal pseudopotential $\hat{V}_{\rm NL}$ is replaced by
$\Psi^{-1}\hat{V}_{\rm NL}\Psi$, where $\Psi$ is a trial wave function.  This
approximation leads to errors that are second order in the error in the trial
wave function.\cite{Mitas_1991} Unfortunately these errors may be of either
sign, undermining the variational principle for the fixed-node DMC
ground-state energy.\cite{Foulkes_2001} Furthermore, divergences in the
localized pseudopotential due to nodes in the trial wave function can result
in instabilities in the DMC algorithm.  The latter two problems can largely be
removed by means of a partial locality approximation known as the ``T-move''
scheme.\cite{Casula_2006,Casula_2010} A more basic issue with the use of
pseudopotentials is that they are generally constructed within the context of
a single-particle method, such as Hartree-Fock theory or density functional
theory (DFT)\@. Only recently has the possibility of developing
pseudopotentials for explicitly correlated many-body methods been
explored.\cite{Trail_2013,Trail_2015}

In this article we will examine a number of practical issues that affect the
use of Dirac-Fock pseudopotentials in QMC calculations.  In particular we will
analyze the widely used Dirac-Fock pseudopotentials introduced by Trail and
Needs, hereafter referred to as TN
pseudopotentials.\cite{Trail_2005a,Trail_2005b} Most QMC calculations for real
materials have featured either first- or second-row atoms.\cite{Needs_2010} An
issue that quickly emerges when one attempts to use TN pseudopotentials for
transition metals ``off the shelf'' in plane-wave DFT orbital-generation
calculations is the presence of ghost states due to the Kleinman-Bylander
representation of the pseudopotentials in the DFT code.\cite{Kleinman_1982} We
demonstrate that this problem can generally be avoided in a straightforward
fashion.  We also investigate the energy variance per electron that can be
achieved in QMC calculations for different elements, and we determine
appropriate plane-wave cutoff energies for each pseudopotential.  We have
investigated various practical issues such as the optimal choice of spherical
integration rule for pseudopotentials in QMC calculations.

The purpose of the present article is not to analyze the accuracy or otherwise
of the TN pseudopotentials, but rather to determine how best to obtain
consistent and precise QMC results using these pseudopotentials in conjunction
with plane-wave basis sets.  This is clearly a necessary first step towards
assessing the accuracy of such calculations.  It is of course also possible to
use localized basis sets, such as Gaussian basis sets or Slater-type basis
sets, in QMC calculations, with or without the use of pseudopotentials.

The rest of this article is arranged as follows.  In
Sec.\ \ref{sec:methodology} we describe the computational methodology that we
have used to analyze the performance of TN pseudopotentials in QMC
calculations.  In Sec.\ \ref{sec:results} we present our results and analyze
their implications for the use of pseudopotentials in QMC calculations.
Finally we draw our conclusions in Sec.\ \ref{sec:conclusions}.

\section{Computational methodology \label{sec:methodology}}

\subsection{TN pseudopotentials}

The TN pseudopotentials are Dirac-Fock average relativistic effective
pseudopotentials optimized for QMC calculations.\cite{Trail_2005a,Trail_2005b}
The core is as large as possible in each case, and pseudopotential data are
provided for the $s$, $p$, and $d$ angular-momentum channels only.  The issue
of the need for higher-angular momentum channels in some cases is discussed in
Ref.\ \onlinecite{Tipton_2014}.  Either $s$, $p$, or $d$ must be chosen to be
the local channel, the potential for which is then applied to all higher
angular-momentum components of the wave function.  By default the $d$
angular-momentum channel is chosen to be local in the TN pseudopotentials.
The other widely used set of Dirac-Fock pseudopotentials for QMC calculations
is that proposed in Refs.\ \onlinecite{Burkatzki_2007} and
\onlinecite{Burkatzki_2008}.  The performance of these two families of
pseudopotentials in studies of small molecules is compared in
Ref.\ \onlinecite{Trail_2014}.

For each element we have taken the first TN pseudopotential listed in the
library at \url{https://vallico.net/casinoqmc/pplib/}.  This is a Dirac-Fock
average relativistic effective pseudopotential tabulated on a radial grid.
Core-polarization corrections\cite{Shirley_1993} are not used in this work.
For many elements, alternative TN pseudopotentials are available.  These are
either constructed to be softer, or constructed using nonrelativistic
Hartree-Fock theory; however, they are not considered in this work.

\subsection{DFT calculations}

The usual starting point for a QMC trial wave function is a Slater determinant
of single-electron orbitals generated using DFT\@.  We have performed
plane-wave DFT calculations in order to determine whether or not each
TN pseudopotential is affected by ghost states, and to generate trial wave
functions for subsequent QMC calculations.

Our DFT orbital-generation calculations were performed using the
\textsc{castep} plane-wave basis code.\cite{Clark_2005} Each atom was placed
in a simple cubic cell of side-length 15 bohr subject to periodic boundary
conditions.  The Perdew-Burke-Ernzerhof (PBE) generalized-gradient
approximation exchange-correlation functional\cite{PBE_1996} was used, and the
plane-wave cutoff energy was 120 Ha in each case (except where the cutoff
energy was varied to investigate convergence).  We then performed
spin-polarized calculations with the appropriate number of up- and down-spin
electrons for the ground-state electronic configuration of the isolated atom.

In order to avoid ghost states we also performed PBE DFT calculations without
a Kleinman-Bylander representation of the semilocal norm-conserving
pseudopotentials.  These calculations were performed with Gaussian basis
functions using the \textsc{molpro} package.\cite{Werner_2012,MOLPRO} Large
Gaussian basis sets were used, with aug-cc-pV5Z Dunning basis
sets\cite{Dunning_1989,Kendall_1992,Peterson_2002} used when available, and
the pseudopotential equivalent used otherwise (for $39 \leq Z \leq 54$ and $72
\leq Z \leq 80$).  All basis sets were uncontracted.  Spin and orbital
constraints were chosen in analog to the \textsc{castep} calculations, with
the total spin constrained to be that of the ground-state configuration, and
up- and down-spin orbitals unrestricted.  For the Gaussian basis calculations
we used the Gaussian expansion of each tabulated TN pseudopotential, available
from the same source as the tabulated pseudopotentials.

Finite-basis errors for the Gaussian basis sets used in this work are expected
to be small, but we have examined the issue in more detail for oxygen and
copper.  The convergence of the total energy with respect to basis-set index
is expected to be rapid and exponential;\cite{Dunning_1989} hence the
basis-set errors in the aug-cc-pV5Z results were estimated by three-point
exponential extrapolation to the complete-basis-set limit using
aug-cc-pV(T,Q,5)Z basis sets.  This provides estimated basis-set errors of
0.002 and 0.005 Ha for oxygen and copper, respectively.

\subsection{Trial wave functions \label{sec:wf}}

The plane-wave orbitals were re-represented in a blip (B-spline)
basis.\cite{Alfe_2004} This is faster to evaluate in a QMC calculation and
allows us to dispense with the unwanted periodic boundary conditions on the
isolated atoms and molecules studied in this work.  For each orbital the grid
of reciprocal-lattice points in the plane-wave expansion was padded out by a
factor of three in each direction before a Fourier transformation to real
space was carried out to obtain the blip representation of the
orbital.\footnote{The grid of reciprocal-lattice points was padded out by a
  factor of two in each direction before Fourier transformation to the
  real-space grid in our tests of the effects of varying the plane-wave cutoff
  energy and the self-consistent-field convergence tolerance.} We verified
that this choice of ``blip grid multiplicity'' is sufficiently fine that the
re-representation of the plane-wave orbitals in a blip basis has no
statistically significant effects on our QMC results.  We optimized three
levels of correlated wave function: Slater-Jastrow (SJ) wave functions with
isotropic electron-electron ($u$) and electron-nucleus ($\chi$) terms in the
Jastrow factor; SJ wave functions with isotropic electron-electron ($u$),
electron-nucleus ($\chi$), and electron-electron-nucleus ($f$) terms in the
Jastrow factor;\cite{Drummond_2004} and Slater-Jastrow-backflow (SJB) wave
functions with isotropic electron-electron ($u$), electron-nucleus $(\chi$),
and electron-electron-nucleus ($f$) terms in the Jastrow factor and isotropic
electron-electron ($\eta$) and electron-nucleus ($\mu$) terms in the backflow
function.\cite{Lopez_2006} In each case we used separate two-electron Jastrow
and backflow terms for parallel- and antiparallel-spin electrons; where there
were different numbers of up- and down-spin electrons we also used different
two-electron Jastrow terms for up-up and down-down pairs.  Likewise, for
spin-polarized atoms, we used separate one-electron Jastrow and backflow terms
for up- and down-spin electrons.  The electron-electron-nucleus Jastrow terms
were of the same form for all combinations of electron spins, however.  For
each atom the $u$, $\chi$, $f$, $\eta$, and $\mu$ functions were smoothly
truncated at interparticle distances of 6, 6, 3, 6, and 6 bohr, respectively.
The only SJB results we report are in our tests of DMC for a small-core copper
atom, in Sec.\ \ref{sec:copper_atom}.

In our variational Monte Carlo (VMC) wave-function optimization calculations
we used the unreweighted variance
minimization\cite{Umrigar_1988,Drummond_2005} and energy
minimization\cite{Nightingale_2001,Toulouse_2007,Umrigar_2007} methods.  We
used more than 10,000 electronic configurations in each optimization.  We also
performed QMC calculations using an uncorrelated product of Slater (S)
determinants for up- and down-spin electrons.

\subsection{DMC calculations \label{sec:dmc_params}}

All our DMC calculations used time steps of $\delta \tau=0.0005$, 0.001,
0.002, and 0.004 Ha$^{-1}$, with the numbers of walkers being inversely
proportional to $\delta \tau$ in each case.  At small time steps the DMC
time-step bias is linear in the time step while the finite-population bias is
inversely proportional to the population;\cite{Umrigar_1993} hence, linearly
extrapolating our DMC energies to zero time step simultaneously removes the
time-step and the population-control biases. We have performed DMC
calculations with and without the T-move scheme\cite{Casula_2006,Casula_2010}
using Slater-Jastrow wave functions with electron-electron ($u$) and
electron-nucleus ($\chi$) terms only in the Jastrow exponent.

\section{Results and discussion \label{sec:results}}

\subsection{Ghost states \label{sec:ghosts}}

Plane-wave DFT codes such as \textsc{castep} typically use the
Kleinman-Bylander\cite{Kleinman_1982} form to represent pseudopotentials for
reasons of efficiency.  Unfortunately, for several elements, recasting the TN
pseudopotentials in Kleinman-Bylander form results in the formation of
so-called ``ghost states'' (spurious low-energy bound
states).\cite{Gonze_1990,Gonze_1991} The Kleinman-Bylander representation is
slightly different to the semilocal representation of a pseudopotential, and
the introduction of ghost states is not simply an artifact of basis-set
incompleteness.  The presence of ghost states gives rise to some or all of the
following symptoms: the failure of the DFT self-consistent-field (SCF) process
to converge; a large difference between the DFT energies obtained with
plane-wave and Gaussian basis sets; the existence of an absurdly low Kohn-Sham
eigenvalue; an absurdly high (unbound) energy when the orbitals are used in
VMC calculations; a very large energy variance; enormous difficulty optimizing
a trial wave function in VMC; and enormous difficulty controlling the
configuration population in a DMC simulation.  Furthermore, these difficulties
may change or disappear when the local channel is changed.

\begin{figure}
\begin{center}
\includegraphics[clip,width=0.45\textwidth]{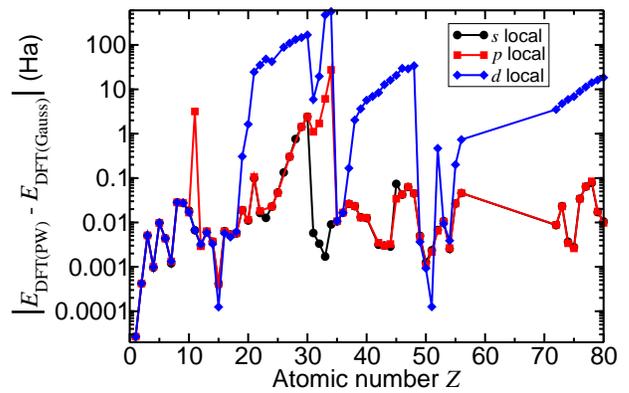}
\caption{(Color online) Magnitude of difference between plane-wave and
  Gaussian DFT total energies $E_{\rm DFT(PW)}$ and $E_{\rm DFT(Gauss)}$ for
  each TN pseudopotential, for different choices of local channel.  No results
  are given for lutetium because of the lack of availability of an accurate
  Gaussian basis set; this does not imply that lutetium is affected by ghost
  states.  The DFT calculations for niobium with $s$ and $p$ local, vanadium,
  iron, and nickel with $p$ local, and manganese with $d$ local persistently
  failed to converge and are not shown in the graph.  \label{fig:EDFT_v_Z}}
\end{center}
\end{figure}

\begin{figure*}
\begin{center}
\includegraphics[clip,width=0.8\textwidth]{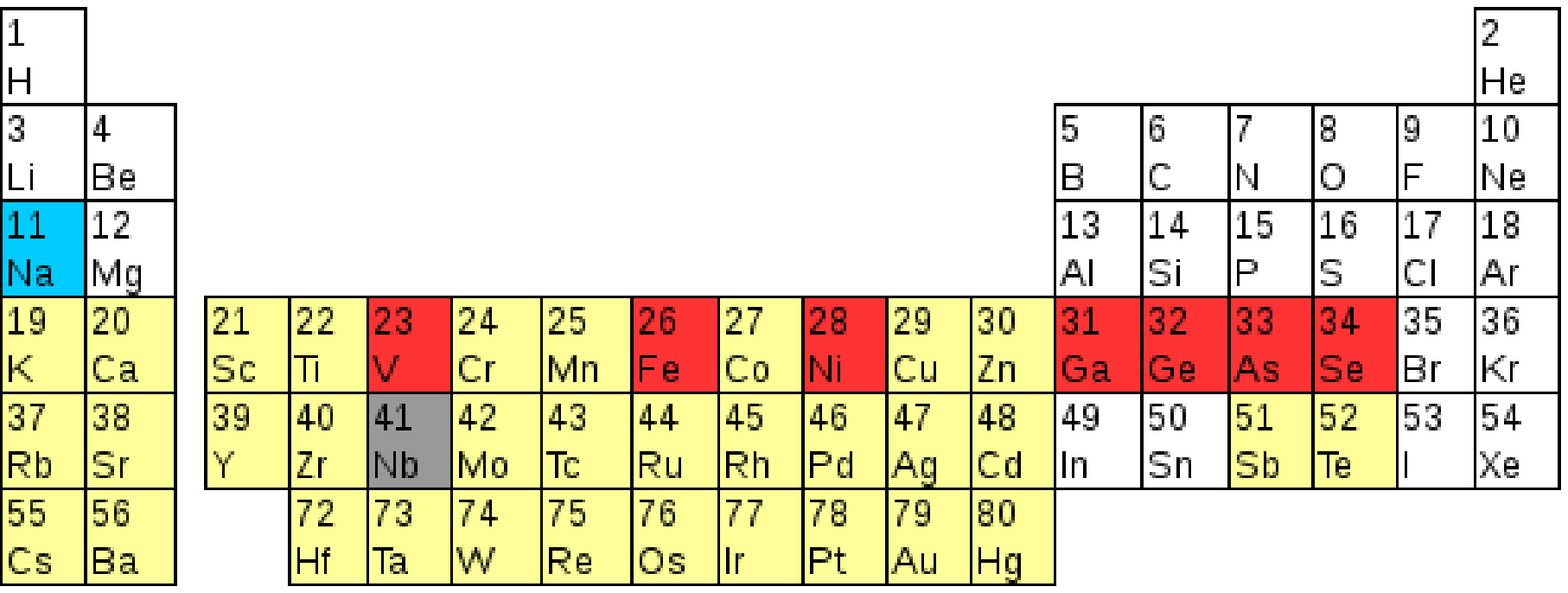}
\caption{(Color online) TN pseudopotentials that are affected by ghost states
  for different choices of local channel in plane-wave DFT calculations.  For
  elements in white cells, there are no ghost states; for elements in blue
  cells, choosing $p$ (but not $s$ or $d$) to be local results in ghost
  states; for elements in yellow cells, choosing $d$ (but not $s$ or $p$) to
  be local results in ghost states; for elements in red cells, choosing $p$ or
  $d$ (but not $s$) to be local leads to ghost states; and elements in grey
  cells are haunted by ghost states for any choice of local channel.  Gaussian
  DFT results are not available for lutetium ($Z=71$), which is therefore
  omitted. \label{fig:periodic_table_dft}}
\end{center}
\end{figure*}

In Fig.\ \ref{fig:EDFT_v_Z} we plot the difference between the DFT energy
obtained using a plane-wave basis with a Kleinman-Bylander representation of
the pseudopotential and the DFT energy obtained using a Gaussian basis for
each TN pseudopotential.  Choosing the $d$ channel to be local often leads to
a relatively enormous difference between the plane-wave and Gaussian DFT
results, strongly suggesting a problem caused by a ghost state.  Choosing the
$s$ channel to be local avoids this problem in every case apart from niobium.
Choosing the $p$ channel to be local avoids the problem with ghost states in
most cases.  In Fig.\ \ref{fig:periodic_table_dft} we indicate which TN
pseudopotentials are likely to be affected by ghost states for different
choices of local channel, based on this analysis of the DFT energies.

The set of Kohn-Sham eigenvalues obtained for the copper TN pseudopotential is
shown in Table \ref{table:copper_eigenvalues}. Results are given for both
Gaussian and plane-wave basis calculations, with the $s$ channel taken as
local for the plane-wave calculations.  In both cases the basis-set errors
were estimated. The Gaussian basis-set eigenvalues were obtained by
three-point exponential extrapolation to the complete-basis-set limit using
aug-cc-pV(T,Q,5)Z basis sets, with errors estimated as the absolute difference
between the extrapolated eigenvalues and those for the aug-cc-pV5Z basis
set. The plane-wave basis-set eigenvalues were obtained by linear
extrapolation with respect to $E_{\rm cut}^{-3/2}$, where $E_{\rm cut}$ is the
plane-wave cutoff energy, using cutoff energies of $240$, $300$, and $360$ Ha.
Errors were estimated as the absolute difference between the extrapolated
values and those obtained for an energy cutoff of $360$ Ha.  As suggested by
Fig.\ \ref{fig:EDFT_v_Z}, the largest error arises for the $d$-orbitals and
plane-wave basis calculations.  Total energies were obtained using the same
extrapolation and error-estimation procedure.  For the Gaussian and plane-wave
basis sets the total energies were $-52.661(5)$ and $-52.49(1)$ Ha,
respectively.  The difference between them is significant, but sufficiently
small that it can be ascribed to the Kleinman-Bylander representation with no
ghost states present.

\begin{table}
\begin{center}
\caption{Kohn-Sham eigenvalues for the occupied orbitals of the copper atom
  represented by a TN pseudopotential.  The basis set is either Gaussian or
  plane-wave; a Kleinman-Bylander representation of the pseudopotential is
  used in the latter case.  Estimated basis-set errors are shown in brackets.
    \label{table:copper_eigenvalues}}
\begin{tabular}{lcr@{.}lr@{.}l}
\hline \hline

& & \multicolumn{4}{c}{Eigenvalues (Ha)} \\

\raisebox{1.5ex}[0pt]{Orbital} & \raisebox{1.5ex}[0pt]{Basis} &
\multicolumn{2}{c}{Spin-up} & \multicolumn{2}{c}{Spin-down} \\

\hline

$4s$ & Gaussian   & $-0$&$158987(4)$ & $-0$&$12136(6)$   \\

$3d$ & Gaussian   & $-0$&$262667(1)$ & $-0$&$258891(1)$  \\

$4s$ & Plane-wave & $-0$&$1582(1)$   & $-0$&$1176(1)$    \\

$3d$ & Plane-wave & $-0$&$2762(1)$   & $-0$&$2715(2)$    \\

\hline \hline
\end{tabular}
\end{center}
\end{table}

The presence of ghost states makes QMC work meaningless or impossible;
however, inexperienced users may wrongly ascribe the problems encountered to
the general difficulty of optimizing QMC trial wave functions.  Eliminating
ghost states from DFT orbital-generation calculations is a necessary but not
sufficient condition for accurate QMC work.  Even if the orbitals generated in
the DFT calculation are unaffected by ghost states, the choice of local
channel may still affect the behavior of the subsequent QMC calculations.

\begin{figure*}
\begin{center}
\includegraphics[clip,width=0.8\textwidth]{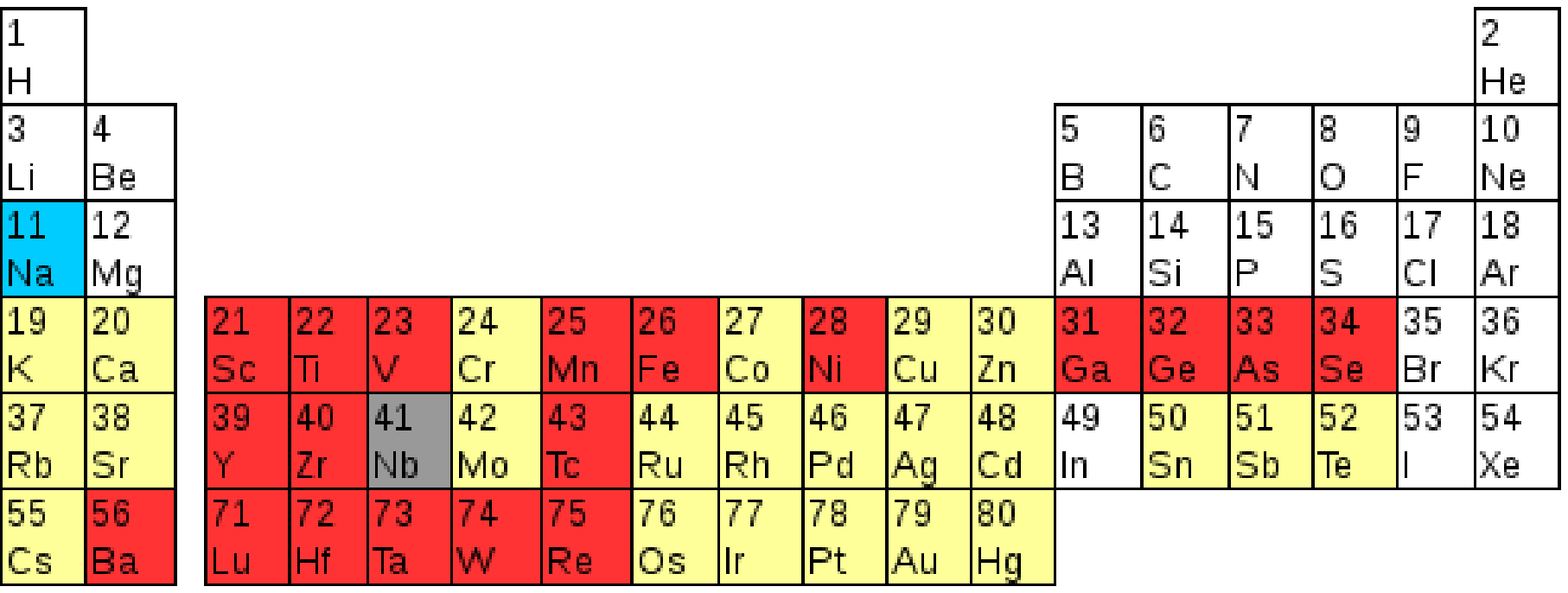}
\caption{(Color online) TN pseudopotentials that are affected by ghost states
  or similar difficulties for different choices of local channel in QMC
  calculations using plane-wave DFT orbitals.  For elements in white cells,
  there are no ghost-state-like symptoms; for elements in blue cells, choosing
  $p$ (but not $s$ or $d$) to be local results in ghost-state-like symptoms;
  for elements in yellow cells, choosing $d$ (but not $s$ or $p$) to be local
  results in ghost-state-like symptoms; for elements in red cells, choosing
  $p$ or $d$ (but not $s$) to be local leads to ghost-state-like symptoms; and
  elements in grey cells are affected by ghost-state-like symptoms for any
  choice of local channel.  \label{fig:periodic_table}}
\end{center}
\end{figure*}

Figure \ref{fig:periodic_table} indicates which TN pseudopotentials are
affected by ghost-state-like symptoms in QMC calculations using a
plane-wave/blip basis for different choices of local channel (used in both the
DFT and the subsequent QMC calculations).  The criteria used for judging that
a particular pseudopotential with a particular choice of local channel is
problematic are the symptoms listed near the beginning of
Sec.\ \ref{sec:ghosts}.  Only one TN pseudopotential, niobium, is apparently
completely unusable in plane-wave calculations.  The problem with niobium is
not currently understood. In every other case, the problem of ghost-state-like
symptoms can be avoided by choosing the $s$ channel to be local in the
plane-wave DFT calculation.

We have verified that in QMC calculations that use orbitals generated with the
$s$ channel chosen to be local, the local channel can either be left as $s$ or
(preferably) changed to $d$; no symptoms of ghost states occur in either case.
By contrast, there are several elements for which choosing the $p$ channel to
be local does not cause any problems in the DFT calculation but does adversely
affect the subsequent QMC calculations, as can be seen by comparing
Figs.\ \ref{fig:periodic_table_dft} and \ref{fig:periodic_table}. Applying the
potential for the $d$ channel to higher angular-momentum components in the QMC
calculation is expected to be more accurate in principle than applying the
potential for the $s$ channel; furthermore, as shown below, choosing $d$ to be
local in QMC reduces the achievable variance in many cases.

Our previous experience with plane-wave DFT calculations for $3d$ transition
metals suggests that choosing $p$ to be local provides reasonable energies,
but is very vulnerable to unstable convergence and/or apparent convergence to
different final energies for different initial conditions. A likely
explanation for this behavior is a ghost state of similar energy to the actual
state. This may well be the cause of the difference between the elements
highlighted as being problematic when $p$ is local in
Figs.\ \ref{fig:periodic_table_dft} and \ref{fig:periodic_table}.  For the
transition metal pseudopotentials the $s$ and $d$ characters of the wave
functions are expected to be dominant near the nuclei, with the $p$ character
being either minimal or zero. This suggests that taking $s$ rather than $p$ to
be local in DFT calculations will be the most accurate choice, as it avoids an
approximate projector representation for this channel.

\subsection{Recommended plane-wave cutoff energies for TN
  pseudopotentials \label{sec:pw_cutoffs}}

We extrapolated the DFT energy of each pseudoatom to basis-set completeness by
assuming the error scales as $E_{\rm cut}^{-3/2}$, where $E_{\rm cut}$ is the
plane-wave cutoff energy (i.e., that the error falls off as the reciprocal of
the number of plane waves).  We used two cutoff values to perform the
extrapolation (240 and 360 Ha, except where SCF convergence problems forced us
to use smaller cutoffs).  Having determined the basis-set-complete energy, we
determined the plane-wave cutoff energy required to achieve a given basis-set
error by linear interpolation in the DFT energy as a function of $E_{\rm
  cut}^{-3/2}$.

\begin{figure}
\begin{center}
\includegraphics[clip,width=0.45\textwidth]{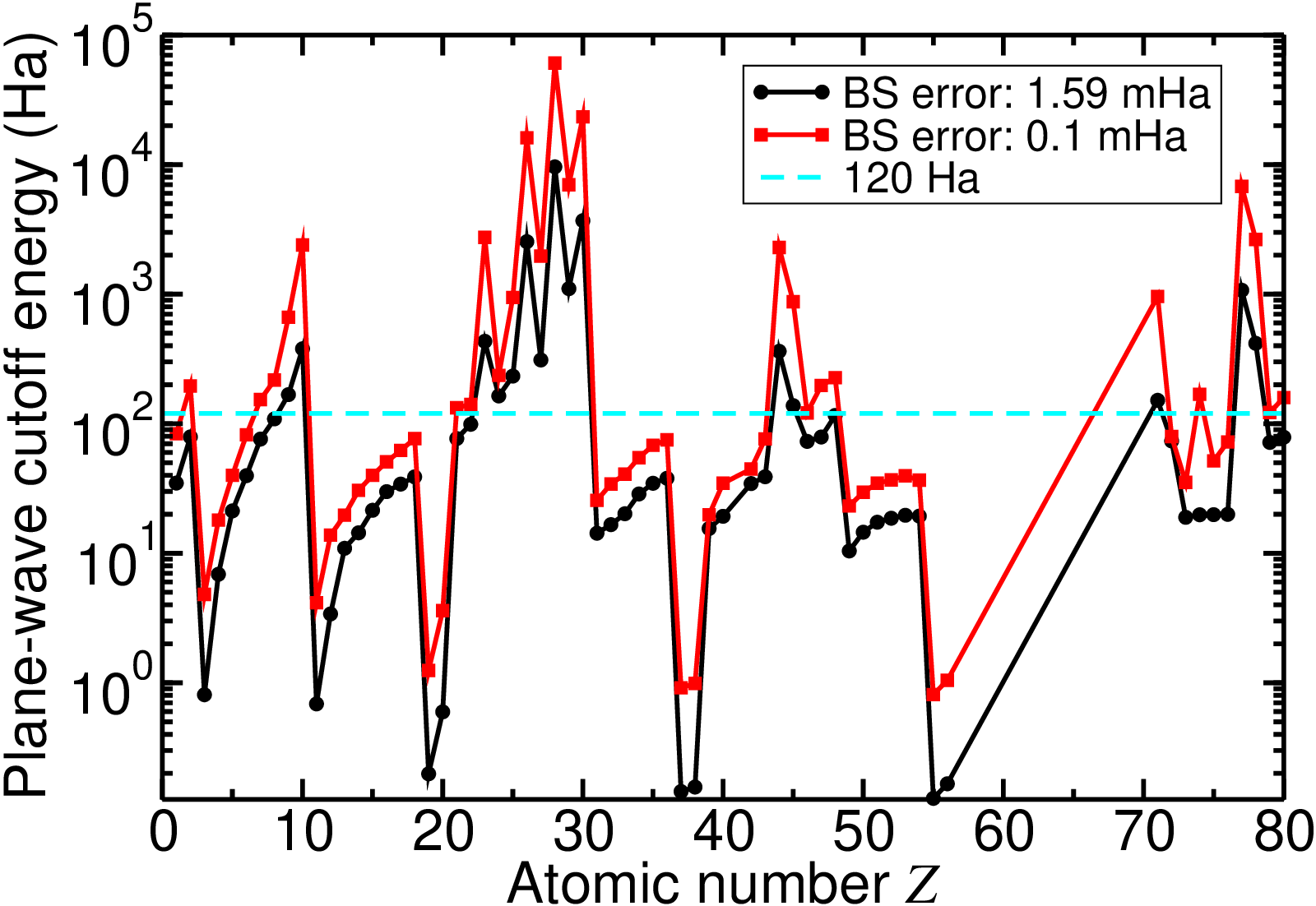}
\caption{(Color online) Plane-wave cutoff energy required to achieve a given
  level of convergence in the DFT energy per atom for each TN pseudopotential.
  The $s$ channel is chosen to be local in each case.  120 Ha is the cutoff
  energy used for the tests reported in
  Fig.\ \ref{fig:var_v_Z}. \label{fig:reqEcut_v_Z_slocal}}
\end{center}
\end{figure}

We have plotted the plane-wave cutoff energy required to achieve a given level
of convergence in the DFT total energy against atomic number in
Fig.\ \ref{fig:reqEcut_v_Z_slocal}.  The cutoff energy required to converge
the total energy of each atom to within so-called chemical accuracy (1 kcal
mol$^{-1}$, which is 1.59 mHa) is shown, as is the cutoff energy required to
converge the total energy to 0.1 mHa, an order of magnitude tighter than
chemical accuracy.  The required cutoff energies for the $3d$ transition
metals are in many cases impractically large; hence any attempt to perform
plane-wave-DFT-QMC calculations using the TN pseudopotentials for those atoms
will inevitably encounter problems associated with large variances, and the
outcome will at best rely on a cancellation of errors. An example of the sort
of difficulties encountered for the $3d$ transition metals is given in
Sec.\ \ref{sec:copper_atom}.

\begin{figure}
\begin{center}
\includegraphics[clip,width=0.45\textwidth]{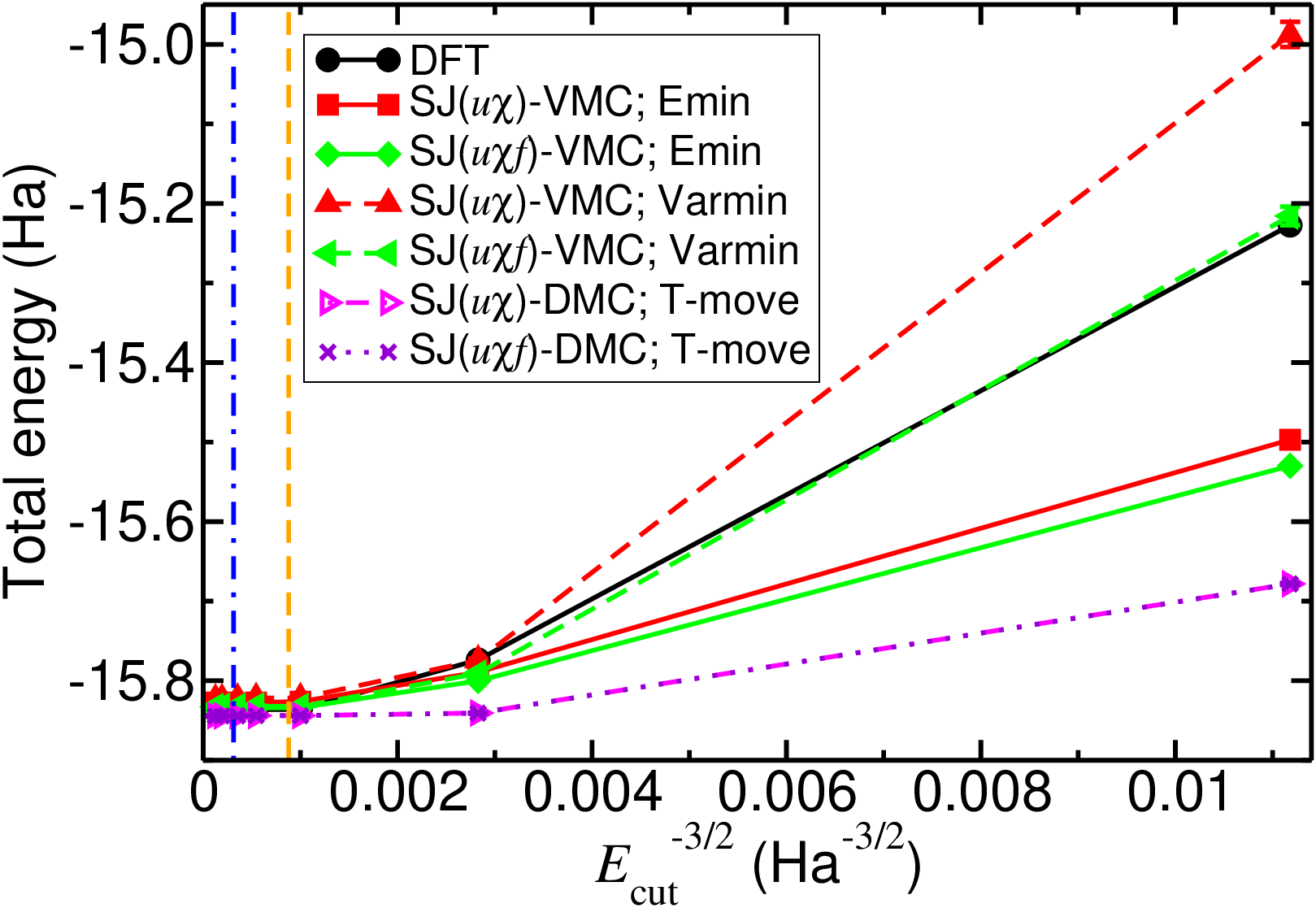}
\\[1em] \includegraphics[clip,width=0.45\textwidth]{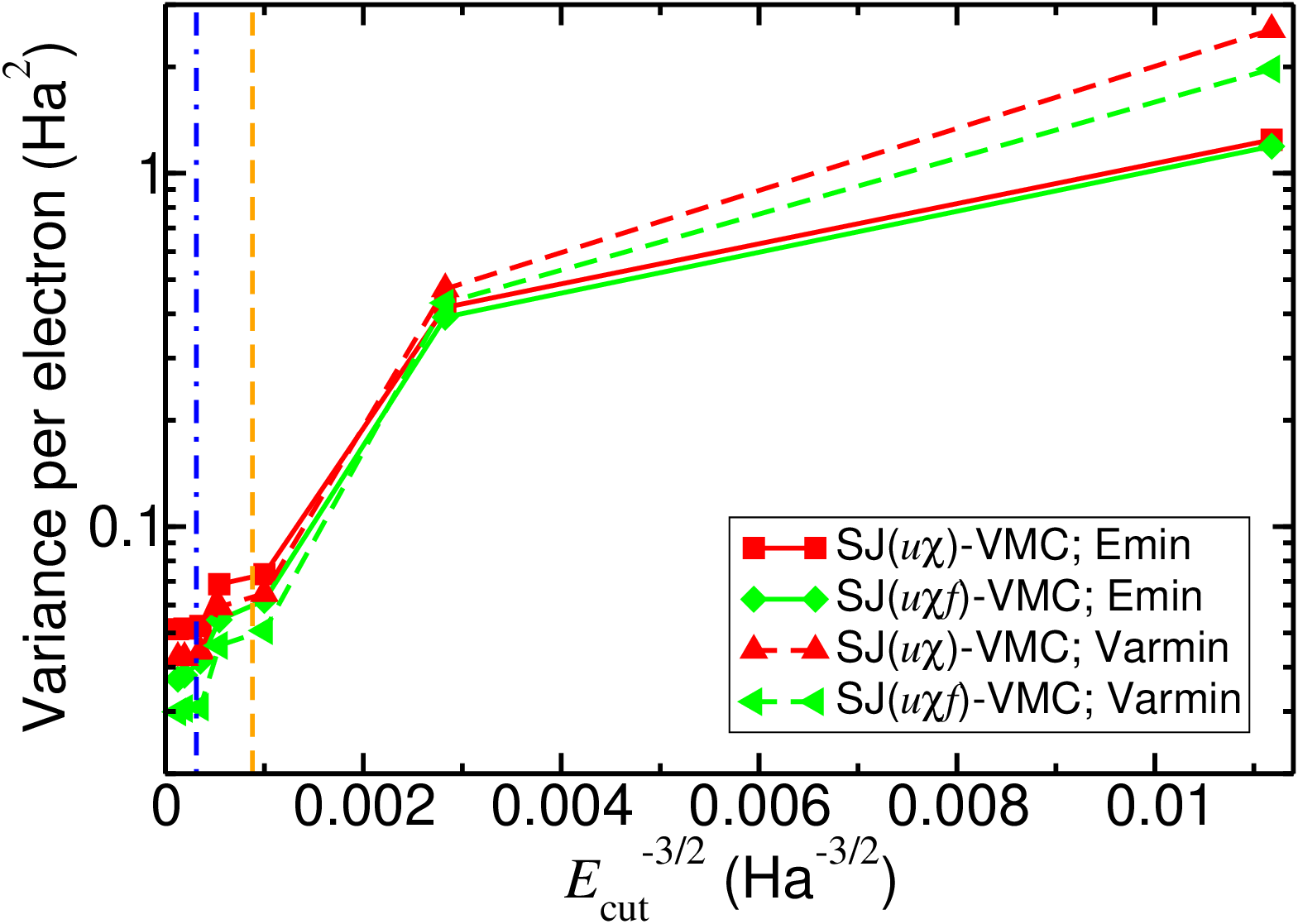}
\caption{(Color online) Top panel: DFT, VMC, and DMC energies as a function of
  plane-wave cutoff energy $E_{\rm cut}$ for an isolated oxygen atom. Bottom
  panel: VMC energy variance against plane-wave cutoff energy for an isolated
  oxygen atom.  Two different levels of correlated wave function were used in
  the VMC calculations, and energy minimization (``Emin'') and unreweighted
  variance minimization (``Varmin'') were used to optimize the wave
  functions. The dashed and dash-dotted vertical lines show the cutoff
  energies at which the DFT total energy is converged to within chemical
  accuracy and to within 0.1 mHa, respectively. The DMC calculations used the
  T-move scheme. Note that unreweighted variance minimization does not
  actually minimize the true variance of the energy; hence higher variances
  can be obtained with unreweighted variance minimization than with energy
  minimization.\cite{Drummond_2005} \label{fig:oxygen_Evar_v_Ecut}}
\end{center}
\end{figure}

\begin{figure}
\begin{center}
\includegraphics[clip,width=0.45\textwidth]{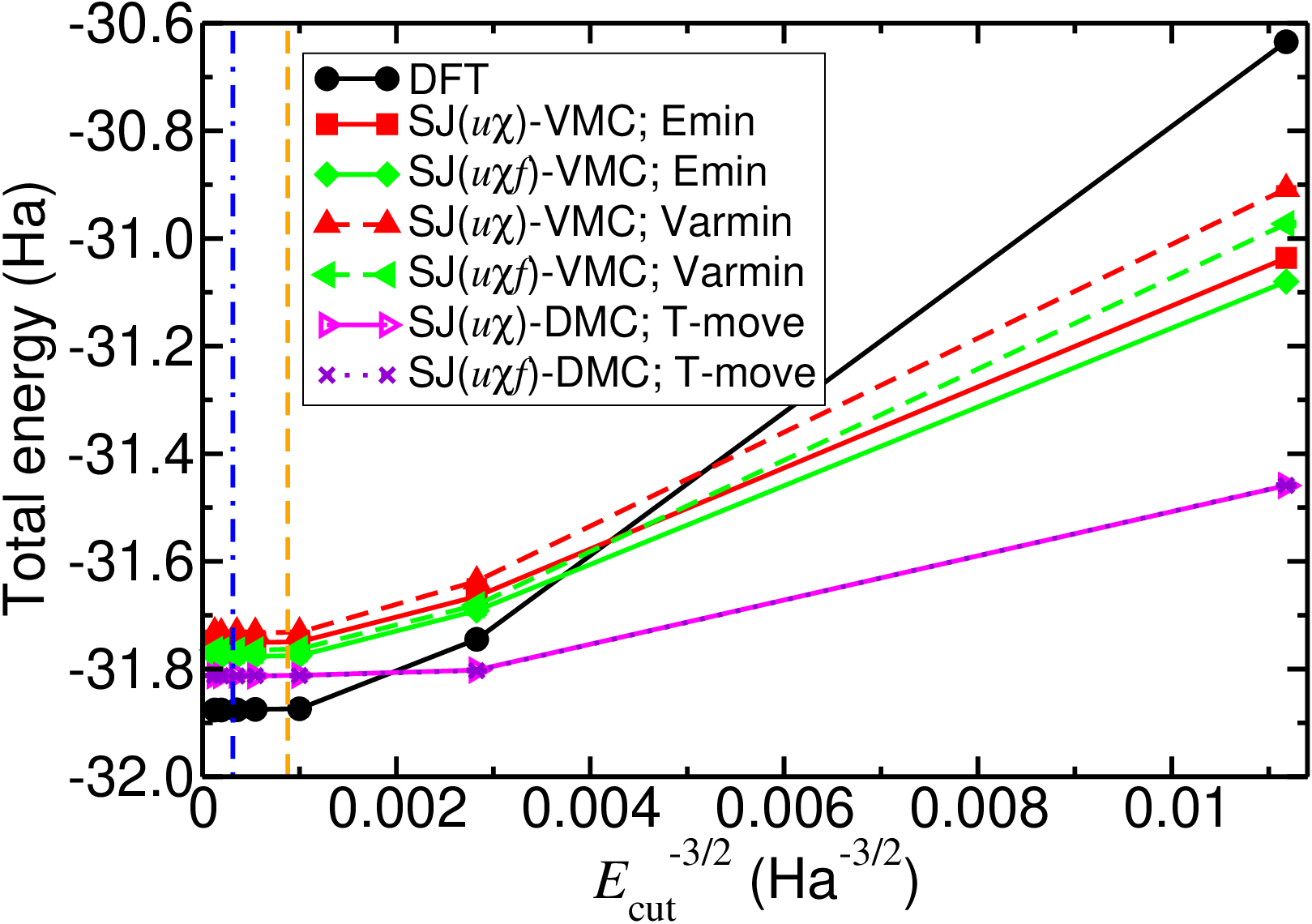}
\\[1em] \includegraphics[clip,width=0.45\textwidth]{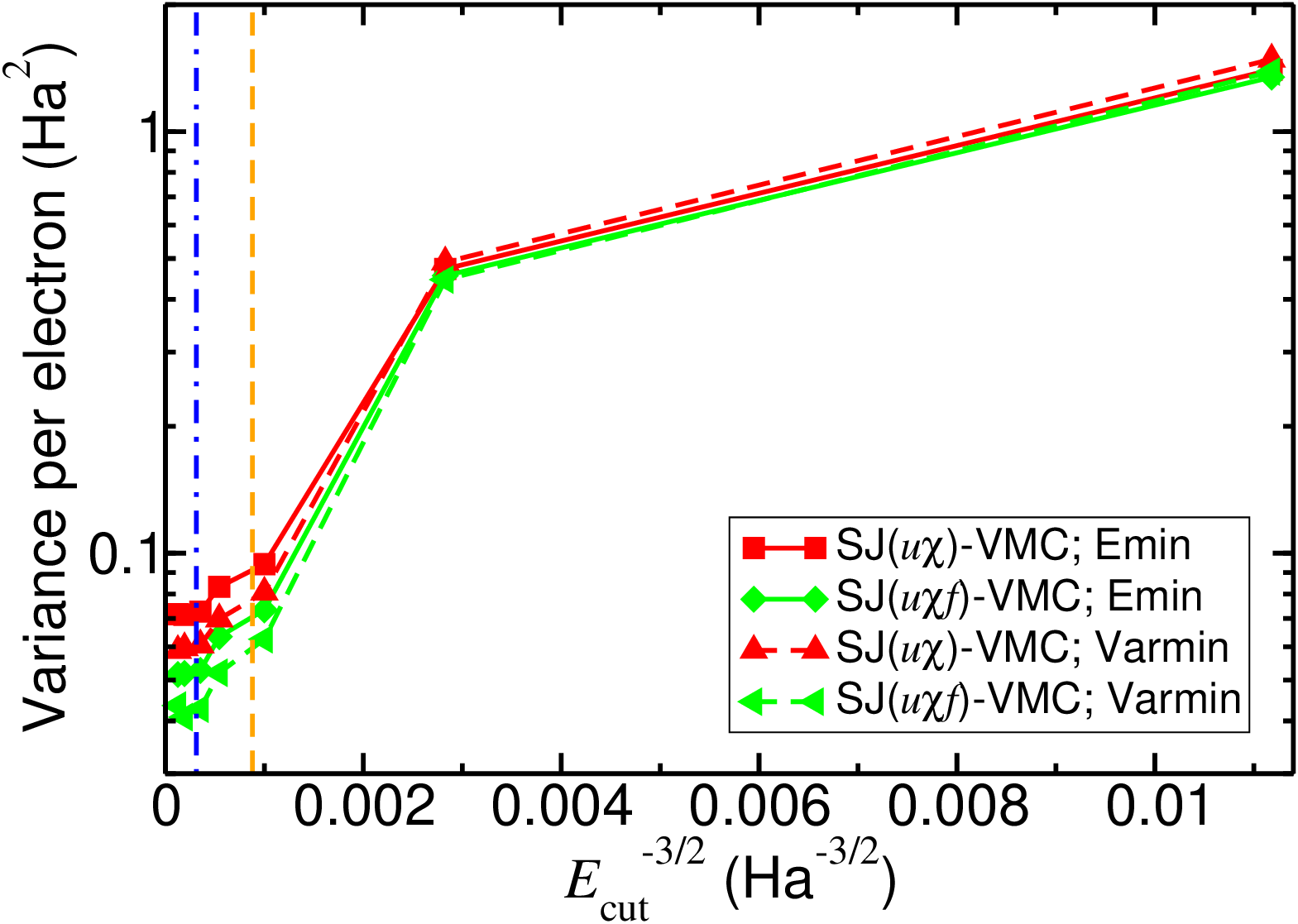}
\caption{(Color online) As Fig.\ \ref{fig:oxygen_Evar_v_Ecut}, but for an
  isolated oxygen molecule. \label{fig:O2_Evar_v_Ecut}}
\end{center}
\end{figure}

The dependence of the DMC energy on the orbitals in the Slater wave function
is in general very weak, because the DMC energy only depends on the trial wave
function via the fixed-node approximation and the pseudopotential locality
approximation; however, the VMC energy, the VMC energy variance, and hence the
efficiency of QMC calculations can depend significantly on the orbitals.  It
has been shown\cite{Ceperley_1986} that, for a given wave-function form, the
efficiency of the importance-sampled DMC algorithm is maximized when the trial
wave function is optimized by energy minimization.  Reducing the finite-basis
error in the DFT total energy per atom provides a better starting point for
optimization of the correlated part of the trial wave function, and the
reduction in the DFT energy with increasing basis-set size translates directly
into a reduction in the VMC energy, as shown for an oxygen atom in
Fig.\ \ref{fig:oxygen_Evar_v_Ecut} and an oxygen molecule in
Fig.\ \ref{fig:O2_Evar_v_Ecut}.  Since QMC calculations are generally intended
to achieve chemical accuracy or higher, it is desirable for the finite-basis
error in the DFT energy to be substantially less than chemical accuracy.
Furthermore, as shown in Figs.\ \ref{fig:oxygen_Evar_v_Ecut} and
\ref{fig:O2_Evar_v_Ecut}, considerable reductions in the VMC variance can be
achieved by increasing the plane-wave cutoff energy up to the value suggested
by the convergence of the DFT total energy to chemical accuracy.  The
performance of unreweighted variance minimization improves significantly once
the basis set becomes adequate.  The results in
Fig.\ \ref{fig:oxygen_Evar_v_Ecut} show that, apart from the lowest plane-wave
cutoff energy studied, the DMC energy is almost independent of the cutoff
energy.

Comparing Figs.\ \ref{fig:oxygen_Evar_v_Ecut} and \ref{fig:O2_Evar_v_Ecut}, it
is clear that there is a significant cancellation of finite-basis errors in
the VMC binding energy of an oxygen molecule, suggesting that one could ``get
away with'' relatively low plane-wave cutoff energies. However, it is more
efficient to use a high plane-wave cutoff energy (such that the DFT total
energy is converged to 0.1 mHa), because it leads to an enormous reduction in
the variance of the energies of the atom and the molecule, and hence gives
smaller statistical error bars in the binding energy.  When the orbitals are
represented in a blip basis in QMC calculations, the cost of the calculation
is only weakly dependent on the plane-wave cutoff energy.  The cost of the DFT
orbital-generation calculation clearly depends on the cutoff energy, but
orbital generation is usually a negligibly small fraction of the computational
expense of a QMC project.

We have investigated the extent to which it is important to use highly
converged plane-wave orbitals in QMC calculations, i.e., whether it is crucial
to use a tight tolerance for self-consistency in the orbital-generation DFT
calculations.  Although by construction the error in the DFT energy is second
order in the error in the plane-wave coefficients, the error in the nodal
surface (and hence the DMC energy) is first order in the error in the
coefficients, albeit with a small prefactor. Similarly, the errors in the VMC
energy and variance are first order in the error in the plane-wave
coefficients.  As can be seen in Table \ref{table:vary_scf_tol}, there is no
evidence of any need for tight SCF tolerances.  That said, since DFT
calculations are generally negligibly cheap compared with DMC calculations,
there is no good reason for not using a tight convergence criterion.

\begin{table}
\squeezetable
\caption{Effect of varying the SCF convergence tolerance in DFT
  orbital-generation calculations on subsequent VMC calculations for an
  isolated oxygen atom.  The VMC calculations used Slater-Jastrow wave
  functions with $u$, $\chi$, and $f$ terms, which were optimized by energy
  minimization.  The plane-wave cutoff energy was 200
  Ha.  \label{table:vary_scf_tol}}
\begin{tabular}{lccr@{.}l}
\hline \hline

SCF tol.\ (Ha) & DFT en.\ (Ha) & VMC en.\ (Ha) & \multicolumn{2}{c}{Var.\ per
  e.\ (Ha$^2$)} \\

\hline

$10^{-6}$ & $-15.8371$ & $-15.8353(4)$ & \hspace{1.5em} $0$&$042$ \\

$10^{-7}$ & $-15.8371$ & $-15.8345(3)$ & $0$&$0392$ \\

$10^{-8}$ & $-15.8371$ & $-15.8348(3)$ & $0$&$0388$ \\

$10^{-9}$ & $-15.8371$ & $-15.8350(3)$ & $0$&$0385$ \\

$10^{-10}$ & $-15.8371$ & $-15.8354(3)$ & $0$&$0382$ \\

$10^{-11}$ & $-15.8371$ & $-15.8353(4)$ & $0$&$042$ \\

$10^{-12}$ & $-15.8371$ & $-15.8355(3)$ & $0$&$0383$ \\

$10^{-13}$ & $-15.8371$ & $-15.8348(3)$ & $0$&$0395$ \\

\hline \hline
\end{tabular}
\end{table}

\subsection{Achievable energy variance with each TN pseudopotential}

\begin{figure}
\begin{center}
\includegraphics[clip,width=0.45\textwidth]{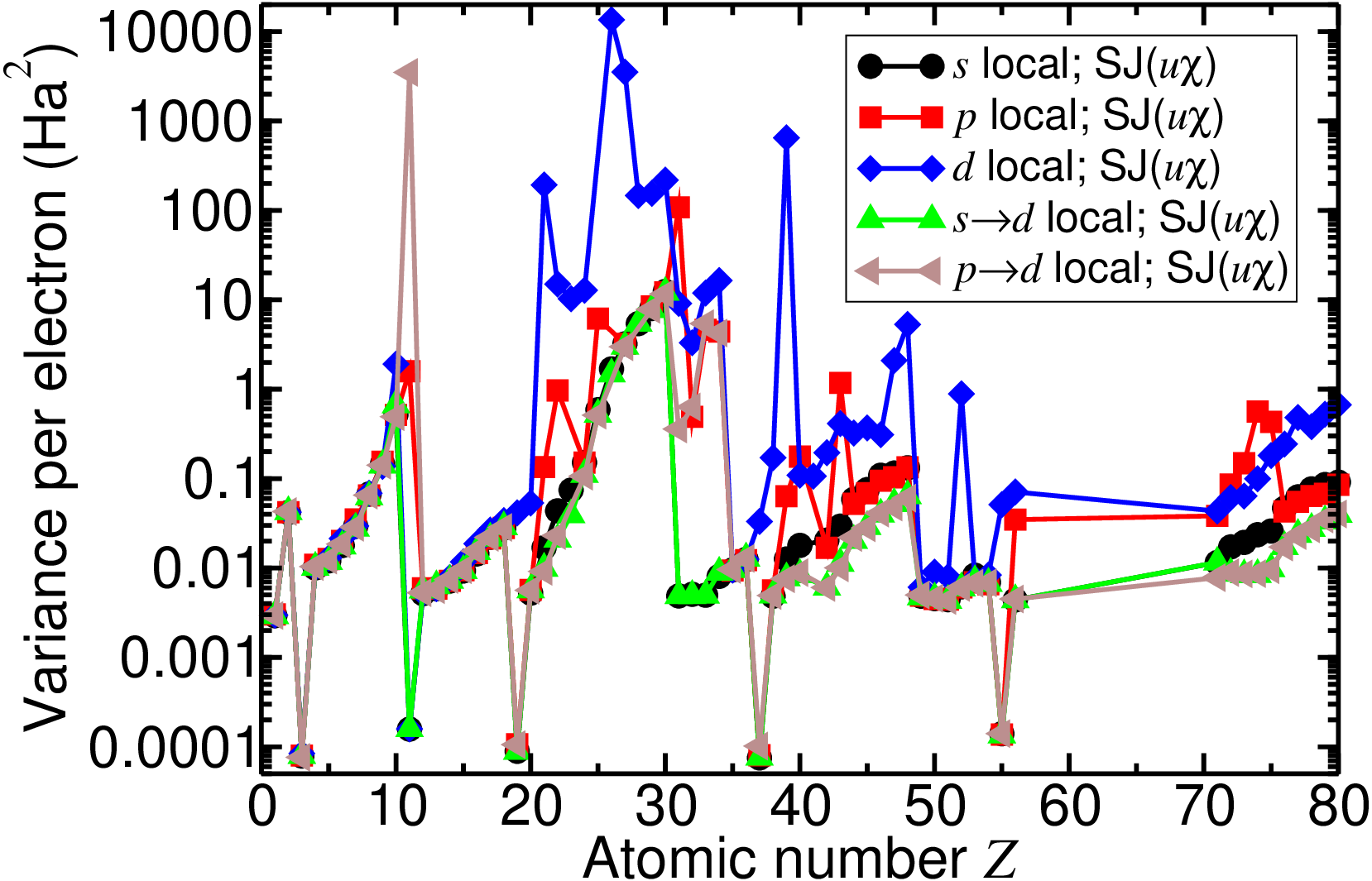}
\\[1em] \includegraphics[clip,width=0.45\textwidth]{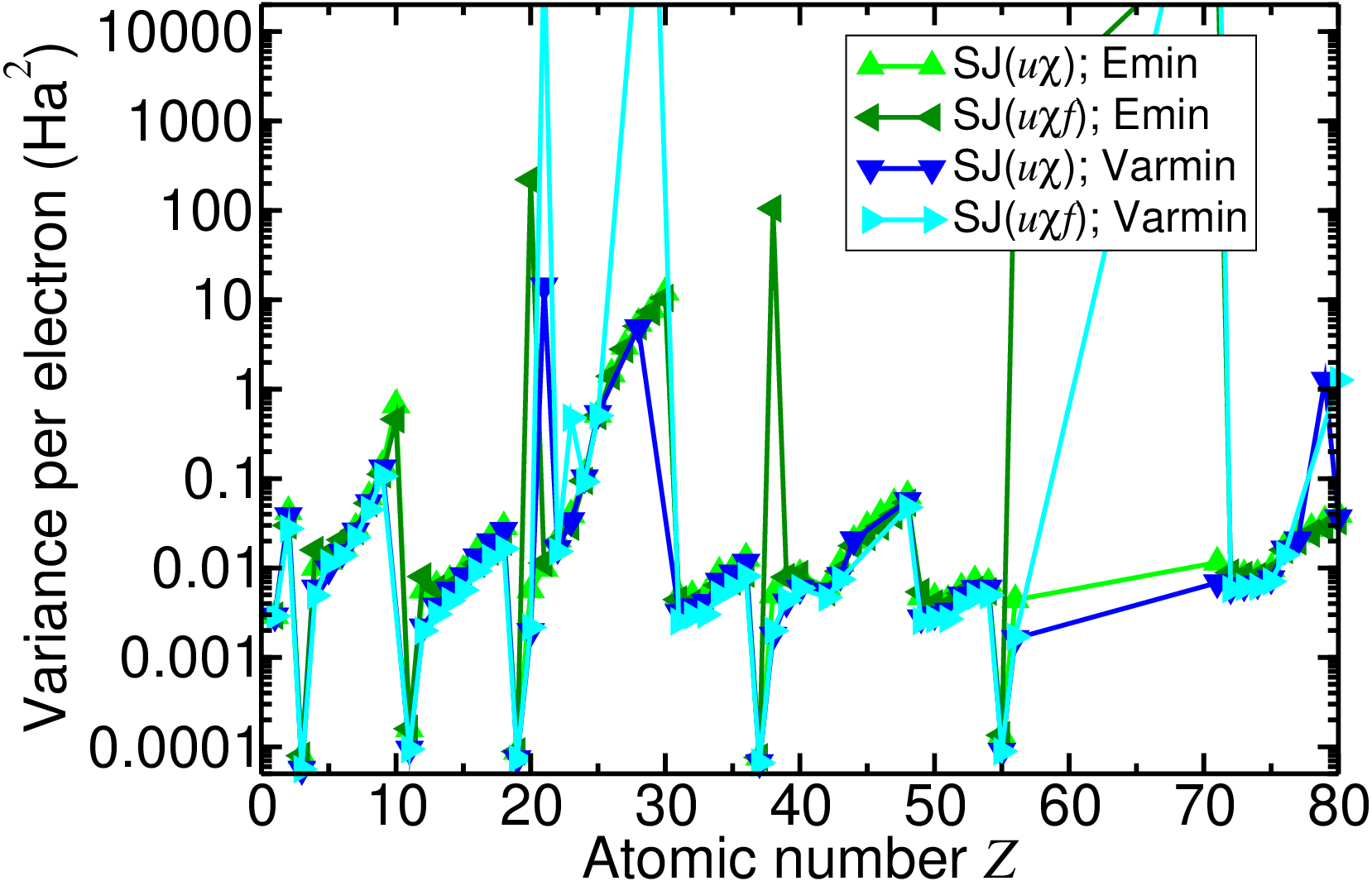}
\caption{(Color online) Top panel: VMC energy variance per electron achieved
  with different angular-momentum channels chosen to be local (``$s\rightarrow
  d$'' indicates that $s$ was chosen to be local in the DFT orbital-generation
  calculation, while $d$ was chosen to be local in the VMC calculation).  We
  used the spin-polarized ground-state electronic configuration in each
  case. Energy minimization was used to optimize the trial wave
  function. Bottom panel: VMC energy variance per electron with the $s$
  channel local in the DFT calculation and the $d$ channel local in the VMC
  calculation, using different levels of Slater-Jastrow trial wave function.
  Energy minimization (``Emin'') and unreweighted variance minimization
  (``Varmin'') were used to optimize the Jastrow factor in each case.
\label{fig:var_v_Z}}
\end{center}
\end{figure}

In Fig.\ \ref{fig:var_v_Z} we plot the VMC energy variance achieved for each
element using a fixed plane-wave cutoff energy of 120 Ha.  Lower variances can
nearly always be achieved by setting $s$ to be the local channel during the
DFT orbital-generation calculation.  The variance is either unchanged or
lowered further if the $d$ channel is subsequently chosen to be local in the
QMC calculation.

Once again it is clear that the most difficult cases by far are the $3d$
transition metals, where the variance is orders of magnitude larger than for
other elements.  Since the results shown in Fig.\ \ref{fig:var_v_Z} were
obtained using a fixed plane-wave cutoff energy of 120 Ha, this is partially
due to the cutoff energy being too small, as shown in
Sec.\ \ref{sec:pw_cutoffs}.  However, for real calculations, it becomes
impractical to use the cutoffs required for these elements.

The lower panel of Fig.\ \ref{fig:var_v_Z} shows that, although unreweighted
variance minimization generally gives a lower variance than energy
minimization, there are more cases in which variance minimization
catastrophically fails.

\subsection{Analysis of DMC total-energy calculations}

\subsubsection{Case study: copper atom \label{sec:copper_atom}}

\begin{figure}
\begin{center}
\includegraphics[clip,width=0.45\textwidth]{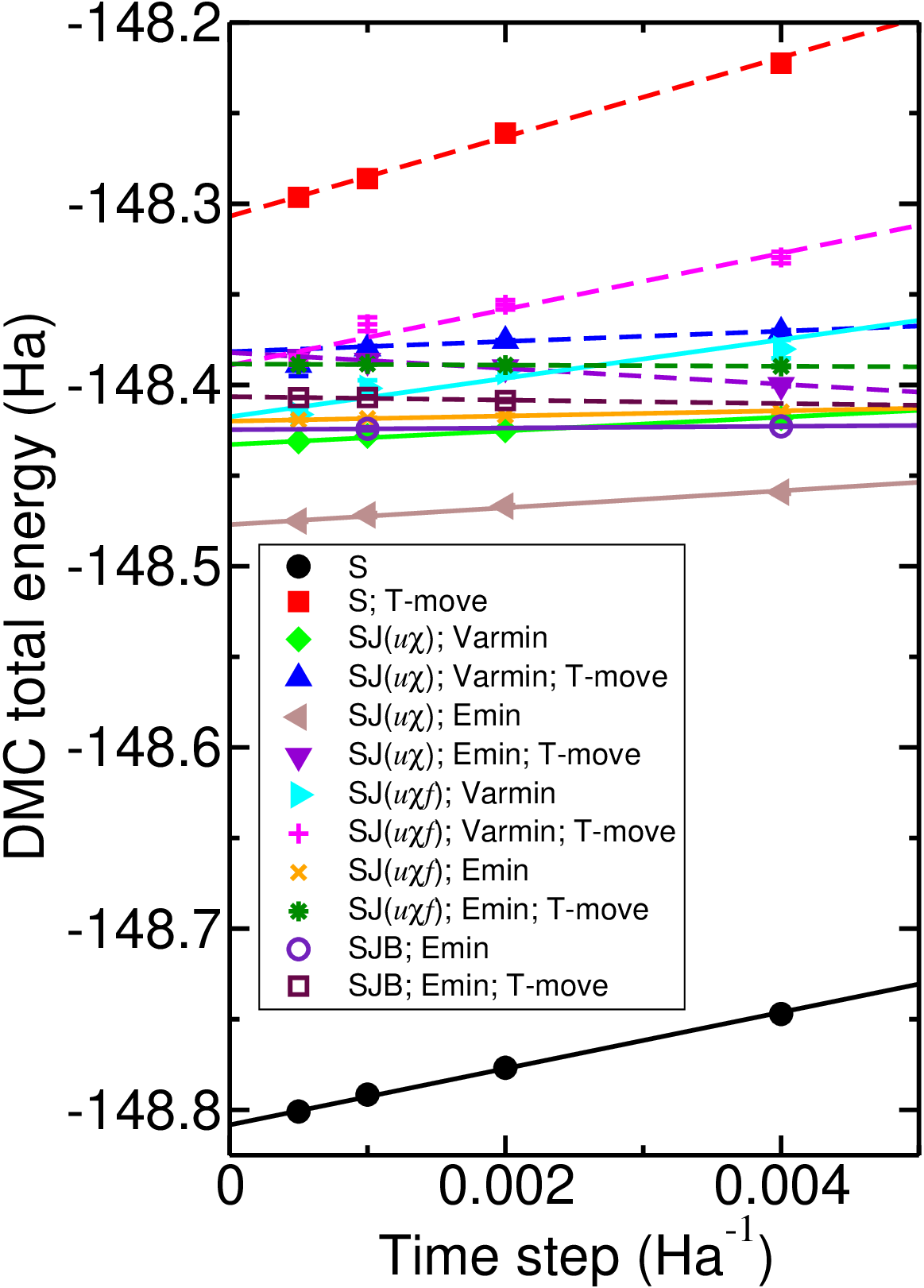}
\caption{(Color online) DMC energy of a small-core copper pseudoatom against
  time step with different trial wave functions, with and without the use of
  T-moves. The trial wave functions were optimized by energy minimization
  (``Emin'') and unreweighted variance minimization (``Varmin'').  The solid
  and dashed lines show fits to the data without and with T-moves,
  respectively.  The orbitals were generated using a plane-wave cutoff energy
  of 120 Ha, and the $s$ channel was chosen to be local to avoid ghost-state
  problems.
\label{fig:copper_atom}}
\end{center}
\end{figure}

To investigate the feasibility of plane-wave-DFT-QMC calculations for $3d$
transition metals, we have carried out a series of calculations for an
isolated copper atom using a small-core pseudopotential, in which the
pseudocharge is 17.  This is not one of the ``standard'' TN pseudopotentials,
but is available via the TN pseudopotential library.  In
Fig.\ \ref{fig:copper_atom} we plot the DMC energy against time step for
various choices of trial wave function and optimization method, with and
without the use of T-moves.

It is clear that the choice of trial wave function and optimization method
affects not only the behavior at finite time step, but also the final DMC
energies extrapolated to zero time step.  In all-electron fixed-node DMC
calculations, the DMC energies at zero time step obtained using Slater and
Slater-Jastrow wave functions are identical, because the nodal surface is not
affected by the Jastrow factor.  However, for copper, pseudopotential locality
errors lead to differences on the scale of several eV between the DMC energies
obtained with different Jastrow factors and without a Jastrow factor.  This
problem is significantly ameliorated by the use of the T-move scheme.

Table \ref{table:copper_atom} shows the DMC energies in
Fig.\ \ref{fig:copper_atom} extrapolated to zero time step and infinite
population, together with the corresponding VMC energies and variances.
Without T-moves the DMC energies depend significantly on the trial wave
function and can be nonvariational.  The spread of DMC energies is
significantly reduced by the T-move scheme, and in particular the spuriously
low energies obtained with poorer wave functions are eliminated.  The change
in the T-move DMC energy resulting from the inclusion of $f$ terms in the
Jastrow factor is small compared with the difference in energy resulting from
the inclusion of backflow.  By contrast, including $f$ terms has a much larger
effect on VMC energies than the inclusion of backflow. This suggests that, if
one has a Slater-Jastrow wave function with $u$, $\chi$, and $f$ terms,
locality errors are small compared with fixed-node errors.  When the wave
function is optimized by energy minimization, the standard error in the DMC
energy is, as expected, significantly lower in general (this is always the
case when T-moves are used).  The energy variances per atom obtained with the
small-core pseudopotential and reported in Table \ref{table:copper_atom} are
between one and two orders of magnitude smaller than the variances per atom
obtained with the standard TN copper pseudopotential and reported in
Fig.\ \ref{fig:var_v_Z}.

\begin{table*}
\caption{DMC total energies in Fig.\ \ref{fig:copper_atom} extrapolated
  linearly to zero time step, together with VMC total energies and VMC energy
  variances per electron for an isolated small-core copper
  pseudoatom. \label{table:copper_atom}}
\begin{tabular}{lcr@{.}lr@{.}lr@{.}lr@{.}l}
\hline \hline

& & \multicolumn{4}{c}{DMC energy (Ha)} & & & & \\

\raisebox{1.5ex}[0pt]{Wave fn.} & \raisebox{1.5ex}[0pt]{Opt.\ meth.} &
\multicolumn{2}{c}{Without T-moves} & \multicolumn{2}{c}{With T-moves} &
\multicolumn{2}{c}{\raisebox{1.5ex}[0pt]{VMC energy (Ha)}} &
\multicolumn{2}{c}{\raisebox{1.5ex}[0pt]{Var.\ per elec.\ (Ha$^2$)}} \\

\hline

Slater &   N/A    & ~~~~$-148$&$808(1)$ & $-148$&$3069(7)$ &
~~~~$-147$&$653(9)$ & \hspace{2em} $1$&$6261$ \\

SJ($u\chi$) & Varmin & $-148$&$4329(4)$ & $-148$&$382(1)$ & $-148$&$158(4)$ &
$0$&$2629$ \\

SJ($u\chi$) & Emin & $-148$&$4770(5)$ & $-148$&$3821(4)$ & $-148$&$196(4)$ &
$0$&$3409$ \\

SJ($u\chi f$) & Varmin & $-148$&$418(3)$ & $-148$&$389(3)$ & $-148$&$032(3)$ &
$0$&$1817$ \\

SJ($u\chi f$) & Emin & $-148$&$4200(2)$ & $-148$&$3884(2)$ & $-148$&$347(3)$ &
$0$&$1837$ \\

SJB & Emin & $-148$&$425(1)$ & $-148$&$4064(4)$ & $-148$&$378(2)$ & $0$&$1167$
\\

\hline \hline
\end{tabular}
\end{table*}

The copper atom is an extreme case that highlights a number of important
issues: pseudopotential locality errors in the DMC total energy can be as
large as several eV per atom, and they manifest themselves by significant
dependence of the DMC energy on the Jastrow factor; pseudopotential locality
errors can be greatly ameliorated by the use of the T-move scheme; if the
plane-wave cutoff energy is smaller than ideal according to
Fig.\ \ref{fig:reqEcut_v_Z_slocal} then energy minimization appears to be more
reliable than unreweighted variance minimization and also results in greater
efficiency in subsequent DMC calculations; and including extra correlation in
the trial wave function using, e.g., electron-electron-nucleus ($f$) terms
reduces both locality errors and time-step bias.

Strategies that prevent numerical problems with copper can be expected to work
for all the other transition metals, because copper has the deepest
$d$ channel.

\subsubsection{Population-control difficulties without T-move scheme}

We have performed a series of DMC calculations for each TN pseudopotential
with and without the T-move scheme, with the $s$ channel chosen to be local.
The simulation parameters are described in Sec.\ \ref{sec:dmc_params}.  When
the T-move scheme was used, every single DMC calculation completed
successfully; however, when the T-move scheme was not used population-control
problems occurred for boron, carbon, oxygen, fluorine, and lutetium.  It is
perhaps surprising that all but one of these problem cases are first-row
atoms.

\subsection{Optimal choice of nonlocal integration grid}

To apply a nonlocal pseudopotential operator to a generic trial wave function
it is necessary to perform integrations over the surface of a sphere in order
to project out the different angular-momentum components of the trial wave
function.  Seven different spherical integration rules are presented in
Ref.\ \onlinecite{Mitas_1991}, and we refer to these as Rules 1--7.  The
number of grid points increases roughly as the square of the rule number, and
each rule integrates an expansion in spherical harmonics $Y_{lm}$ up to
$l=l_{\rm max}$ exactly, where $l_{\rm max}$ increases approximately linearly
with rule number.  In a QMC calculation the spherical grid is typically
rotated randomly before each numerical integration, so that the numerical
integrals are unbiased random estimates of the angular-momentum components of
the trial wave function.  Using an integration scheme with a small number of
grid points has the obvious advantage of reducing the cost of each
local-energy evaluation in a QMC calculation; on the other hand, using a
smaller number of grid points increases the random error in the integration,
and hence the standard error in the mean energy.  It is therefore expected
that there is an optimal numerical integration rule.

Two examples of the effects of using different nonlocal integration rules are
reported in Tables \ref{table:nlgrid_o2} and \ref{table:nlgrid_inse}, for an
oxygen molecule and a monolayer of hexagonal indium selenide, respectively.
For the oxygen molecule the most efficient integration rule is Rule 3, which
uses six points and would be exact for an expansion in spherical harmonics up
to $l_{\rm max}=3$. For indium selenide the optimal rule is Rule 2.
However, since the efficiency falls off very much more steeply when the
integration rule is too small than when it is too large, we recommend that
Rule 4 be chosen to be the default.  Rule 4 uses 12 points and would be exact
for an expansion in spherical harmonics up to $l_{\rm max}=5$. Nevertheless,
QMC practitioners should be aware of the possibility of substantially
increasing the efficiency of their calculations by choosing Rule 3 instead of
Rule 4.

\begin{table}
\caption{VMC total energy, variance per electron, and efficiency with
  different nonlocal pseudopotential integration rules for an isolated O$_2$
  molecule.  The efficiency is defined as the reciprocal of the product of the
  average walltime per iteration, the variance, and the mean decorrelation
  period.  The calculations used a plane-wave cutoff energy of 120 Ha and a
  Slater-Jastrow trial wave function with electron-electron ($u$),
  electron-nucleus ($\chi$), and electron-electron-nucleus ($f$) terms, which
  was optimized by unreweighted variance
  minimization. \label{table:nlgrid_o2}}
\begin{tabular}{lr@{}lr@{}lr@{}l}
\hline \hline

Rule & \multicolumn{2}{c}{VMC en.\ (Ha)} &
\multicolumn{2}{c}{Var.\ per e.\ (Ha$^2$)} & \multicolumn{2}{c}{Effic.\
  (Ha$^{-2}$s$^{-1}$)} \\

\hline

1 & ~~~$-31$&$.74(1)$   & \hspace{2em} $86$&$.4$  & \hspace{3.5em} $7$& \\

2 & $-31$&$.765(5)$  & $12$&$.3$   & $49$& \\

3 & $-31$&$.7646(5)$ & $0$&$.077$ & $5050$& \\

4 & $-31$&$.7649(4)$ & $0$&$.059$  & $4510$& \\

5 & $-31$&$.7649(4)$ & $0$&$.059$  & $3660$& \\

6 & $-31$&$.7649(4)$ & $0$&$.059$  & $2520$& \\

7 & $-31$&$.7649(4)$ & $0$&$.059$  & $1770$& \\

\hline \hline
\end{tabular}
\end{table}

\begin{table}
\caption{As Table \ref{table:nlgrid_o2} but for a $3 \times 3$ simulation
  supercell of monolayer hexagonal indium selenide subject to periodic
  boundary conditions. \label{table:nlgrid_inse}}
\begin{tabular}{lr@{}lr@{}lr@{}l}
\hline \hline

Rule & \multicolumn{2}{c}{VMC en.\ (Ha)} &
\multicolumn{2}{c}{Var.\ per e.\ (Ha$^2$)} & \multicolumn{2}{c}{Effic.\
  (Ha$^{-2}$s$^{-1}$)} \\

\hline

1 & ~~$-22$&$.852(9)$ & \hspace{2em} $1$&$.03$  & \hspace{2.5em} $0$&$.09$ \\

2 & $-22$&$.870(1)$ & $0$&$.0152$ & $3$&$.91$ \\

3 & $-22$&$.869(1)$ & $0$&$.0120$ & $2$&$.80$ \\

4 & $-22$&$.870(1)$ & $0$&$.0115$ & $1$&$.95$ \\

5 & $-22$&$.870(1)$ & $0$&$.0115$ & $2$&$.12$ \\

6 & $-22$&$.870(1)$ & $0$&$.0115$ & $1$&$.62$ \\

7 & $-22$&$.870(1)$ & $0$&$.0115$ & $0$&$.94$ \\

\hline \hline
\end{tabular}
\end{table}

\section{Conclusions \label{sec:conclusions}}

We have investigated various practical issues relating to the use of TN
pseudopotentials in QMC calculations.  In particular, we have determined the
plane-wave cutoff energy required to converge the DFT energy per atom to a
given level of accuracy for each atom in the periodic table up to mercury
(with the exception of the lanthanides).  We have shown that ghost states
arising from the Kleinman-Bylander representation of pseudopotentials in
plane-wave DFT calculations are a significant problem when the $p$ or $d$
channel is chosen to be local.  We have shown that, in all cases apart from
niobium, this problem can be avoided by choosing the $s$ channel to be local
in the DFT calculation and the $d$ channel to be local in the subsequent QMC
calculations.

We have investigated the achievable VMC energy variance with a Slater-Jastrow
wave function for each TN pseudopotential.  Our calculations demonstrate that
accurate QMC calculations run off the back of plane-wave-pseudopotential DFT
calculations for $3d$ elements, especially zinc and copper, cannot easily be
achieved with TN pseudopotentials.  The problematically large plane-wave basis
sets required for the $3d$ transition-metal atoms seem to be a direct
consequence of the norm-conserving pseudopotential generation process, so may
not be easily controlled by modifying this generation process.  We have shown
that the T-move scheme significantly increases the reliability of DMC, and we
recommend its use in general.

\begin{acknowledgments}
J.R.\ Trail and R.J.\ Needs acknowledge financial support from the Engineering
and Physical Sciences Research Council (EPSRC) of the U.K.\ via the Critical
Mass Grant [EP/J017639/1].  Computational resources were provided by Lancaster
University's High-End Computing facility.  Supporting research data may be
freely accessed at
\url{http://dx.doi.org/10.17635/lancaster/researchdata/106}.  We acknowledge
useful conversations with Mike Towler and Katharina Doblhoff-Dier.
\end{acknowledgments}

\bibliography{psp_qmc}

\end{document}